\documentclass{ws-ijmpd}
\usepackage[super,compress]{cite}
\usepackage{graphicx}% include figure files
\usepackage{dcolumn}% align table columns on decimal point
\usepackage{bm}% bold math
\usepackage{esdiff}% for derivatives
\usepackage{hyperref}% add hypertext capabilities
\usepackage{placeins}
\usepackage{amsmath}
\usepackage{amssymb}
\graphicspath{{./}{figures/}}% where to find the figures

\begin{document}

\markboth{Dennis Philipp and Volker Perlick}
{Schwarzschild radial perturbations in EF
and PG coordinates}

\catchline{}{}{}{}{}
%
%%%%%%%%%%%%%%%%%%%%%%%%%%%%%%%%%%%%%%%%%%%%%%%%%%%%%%%%%%%%%%%%%%%%

\title{Schwarzschild radial perturbations in Eddington-Finkelstein
and Painlev{\'e}-Gullstrand coordinates}

\author{DENNIS PHILIPP}
\author{VOLKER PERLICK}

\address{ZARM, University of Bremen, 28359 Bremen, Germany\\
dennis.philipp@zarm.uni-bremen.de\\
volker.perlick@zarm.uni-bremen.de}

\maketitle

\begin{history}
\received{28 March 2015}
\accepted{14 April 2015}
\end{history}

\begin{abstract}
In a previous paper we have considered the Regge-Wheeler
equation for fields of spin $s=0$, 1 or 2 on the Schwarzschild
spacetime in coordinates that are regular at the horizon. In
particular, we have constructed in Eddington-Finkelstein 
coordinates exact solutions in terms of series that are
regular at the horizon and converge on the entire open domain 
from the central singularity to infinity.
Here we extend this earlier work in two different directions.
Firstly, we consider in Eddington--Finkelstein coordinates
a \emph{massive} scalar field that can serve as a dark matter 
candidate. Secondly, we extend the treatment of the massless 
case to Painlev\'e--Gullstrand coordinates, which are 
associated with radially infalling observers. 
\end{abstract}

%\keywords{Keyword1; keyword2; keyword3.}

\ccode{PACS numbers: 04.70.Bw, 03.65.Pm, 04.30.Nk, 11.80.Et}

%\tableofcontents

\section{\label{sec:intro} Introduction}
Under the assumption that the back-reaction to the spacetime geometry 
can be neglected, the propagation of scalar, electromagnetic and gravitational perturbations is described by linear wave equations that have to be solved on a given background spacetime. For this article the background is chosen to be the Schwarzschild spacetime. The investigation of wave propagation in such a black-hole spacetime gives insight into typical phenomena like interference patterns and scattering, (quasi-) normal modes and black hole evaporation. In a recent article \cite{PhilippPerlick2015} we have shown how coordinate systems that are regular at the horizon can be used for constructing analytic solutions to the (massless) Regge-Wheeler equation. In particular, we have used Eddington-Finkelstein coordinates for constructing such solutions in terms of series that are convergent on the entire open interval
$r \in \; ]0, \infty [ \;$. We will use these results here and extend the treatment by i) considering additionally a \emph{massive} scalar field, which brings a new parameter into the game, and ii) constructing solutions for the perturbation equations in the observer-associated Painlev\'{e}-Gullstrand coordinates. 

Considering a massive scalar field establishes a connection to cosmology since it may serve as a dark-matter model \cite{Barranco2011}. We have chosen to present this treatment in Eddington-Finkelstein coordinates, because it is the best adapted coordinate system for this situation and allows to penetrate the horizon \cite{PhilippPerlick2015} following the ingoing radiation. Giving the solutions to the general case of a massless bosonic field with spin $s$ on the other hand allows to investigate black hole physics in coordinates of a radially infalling observer and is therefore, at least in principle, connected to a real physical measurement situation.

The article is organized as follows: We start with some preparations in section \ref{sec:prep} where, after introducing our unit conventions, we recall the basic features of two different coordinate systems for the Schwarzschild spacetime, namely Eddington-Finkelstein and Painlev\'{e}-Gullstrand coordinates. Furthermore, the confluent Heun equation that will be the central differential equation in this work is introduced and  some of its properties are briefly discussed.

In section \ref{sec:massive} we consider a massive scalar field on a Schwarzschild background spacetime and give the wave equation that describes the propagation of this field in Eddington-Finkelstein coordinates. We reduce this wave equation to a purely radial differential equation and show how this radial equation can be transformed to a confluent Heun equation. Thereupon, we give the generalized Riemann scheme and the local solutions at the horizon in terms of the standard confluent Heun function $\mathrm{HeunC}$.

In section \ref{sec:PG} we consider a general massless bosonic spin $s$ field in Painlev\'{e}-Gullstrand coordinates that are associated with observers which follow radial timelike geodesics. Starting from the wave equation in this coordinate system we arrive at a purely radial equation that belongs to the class of confluent Heun equations as well. We construct local solutions at the horizon in terms of the confluent Heun function and asymptotic series at spatial infinity. These local solutions have convergent representations only in a bounded domain of convergence, but we examine an analytic continuation procedure to overcome this problem.
\section{\label{sec:prep} Preparation}
\subsection{Eddington-Finkelstein and Painlev{\'e}-Gullstrand coordinates}
For the present article we will use geometrical units such that the speed of light $c$ and Newton's gravitational constant $G$ are set to unity. On top of that, it is convenient for the following analysis to measure all radial distances in units of the Schwarzschild radius $r_s = 2M$. With this convention the horizon of the Schwarzschild black hole is located at $r=1$ and the light-sphere, where circular lightlike orbits exist, is found at $r=3/2$. The Schwarzschild metric, written in the usual Schwarzschild coordinates, becomes then
\begin{align}
g_{\mu\nu} dx^\mu dx^\nu = - \dfrac{(r-1)}{r} dt^2 + \dfrac{r}{(r-1)} dr^2 + r^2 \left( d\vartheta^2 + \sin^2 \vartheta d \varphi^2 \right) \, . \label{eq:Schwarzschild_metric}
\end{align}
As an alternative to the usual Schwarzschild coordinates $(t,r,\vartheta,\varphi)$ we can introduce other coordinate systems to describe the same spacetime geometry. Two examples that we employ in this work are the Eddington-Finkelstein (EF) and the Painlev\'{e}-Gullstrand (PG) coordinates which are both regular at the horizon. For details on regular coordinate systems for the Schwarzschild spacetime we refer the reader to Ref. \refcite{MartelPoisson2001} and references therein.
The ingoing version of the EF coordinates\footnote{In 1924 Eddington \cite{Eddington1924} was the first to discover this coordinate system which demontrates that the singularity at the horizon is a coordinate artifact and not a physical singularity, even though he never stated this explicitly. The same coodinate system was rediscovered by D. Finkelstein in 1958 \cite{Finkelstein1958}.} is obtained using
\begin{subequations}
\begin{align}
v_{EF} &= t+r+\log(r-1) \, , \\
u &\equiv r \, ,
\end{align}
\end{subequations}
and the Eddington-Finkelstein coordinate $v$ is constant along all radially ingoing lightlike geodesics. The Painlev\'{e}-Gullstrand coordinates \footnote{P. Painlev\'{e} (1921) and A. Gullstrand (1922) independently discovered a spherically symmetric metric without realizing that it was just the Schwarzschild metric in new coordinates.} are in our units given by
\begin{subequations}
\begin{align}
v_{PG} &= t + 2\sqrt{r} + \log \left| \dfrac{\sqrt{r}-1}{\sqrt{r}+1} \right| \, ,\\
u &\equiv r \, ,
\end{align}
\end{subequations}
and are associated with infalling observers that follow radial timelike geodesics starting from spatial infinity \cite{MartelPoisson2001}. Both coordinate systems are obtained by a coordinate transformation starting with usual Schwarzschild coordinates $t$ and $r$, while the angular coordinates are not changed to preserve the symmetry of the situation. The Schwarzschild metric is not diagonal anymore in these coordinate systems but the apparent singularity at the horizon can be characterized as being only a coordinate singularity since the metric is invertible at $r=1$ in EF and PG coordinates.
In the following, we will omit the subscripts ``$EF$'' and ``$PG$'' at the new time coordinate $v$ since the two distinct cases are treated in separate sections.
\subsection{The confluent Heun equation}
The singly confluent Heun equation (CHE), which belongs to Heun's class of differential equations \cite{Heun1888, RonveauxBook, SlavyanovLayBook, MapleHeun}, can be given in the canonical form
\begin{align}
\left[ \diff[2]{}{z} + \left( a + \dfrac{b+1}{z} + \dfrac{c+1}{z-1} \right) \diff{}{z} + \left( \dfrac{\mu}{z} + \dfrac{\nu}{z-1} \right) \right] y(z) = 0 \, , \label{eq:CHE_MapleForm}
\end{align}
where the two parameters $d$ and $e$ are defined in terms of all the other ones by
\begin{subequations}
\begin{align}
\mu &= \dfrac{1}{2} \left( a-b-c-2e+ab-bc \right) \, ,\\
\nu &= \dfrac{1}{2} \left( a+b+c+2d+2e+ac+bc \right) \, .
\end{align}
\end{subequations}
The form \eqref{eq:CHE_MapleForm} is also called the Maple-form of the CHE since it is implemented in the modern computer algebra system Maple \cite{MapleHeun}. The CHE \eqref{eq:CHE_MapleForm} possesses regular singularities at $z=0,1$ and a single irregular singular point at $z=\infty$ \cite{ SlavyanovLayBook,RonveauxBook,MapleHeun}.

At the regular singularities local solutions can be obtained in terms of Frobenius series \cite{Frobenius1873} or be given using the standard confluent Heun function HeunC that is implemented in Maple as well. For details about the definition of this confluent Heun function the reader is referred to, e.g., Refs. \refcite{PhilippPerlick2015,MapleHeun}.
At the irregular singularity asymptotic series, also called Thom\'{e}-type solutions, can be constructed \cite{SlavyanovLayBook}. The behavior of the local solutions depends on the characteristic exponents of the considered singularity. For details on the construction of these local solutions, that involves the so-called generalized Riemann scheme (GRS) as a very useful tool, the reader is referred to Ref.  \refcite{PhilippPerlick2015} and the excellent book on singular differential equations \refcite{SlavyanovLayBook}.
\section{\label{sec:massive} Massive scalar field in EF coordinates}
\begin{figure}[t]
\begin{center}
\includegraphics[width=0.49\textwidth]{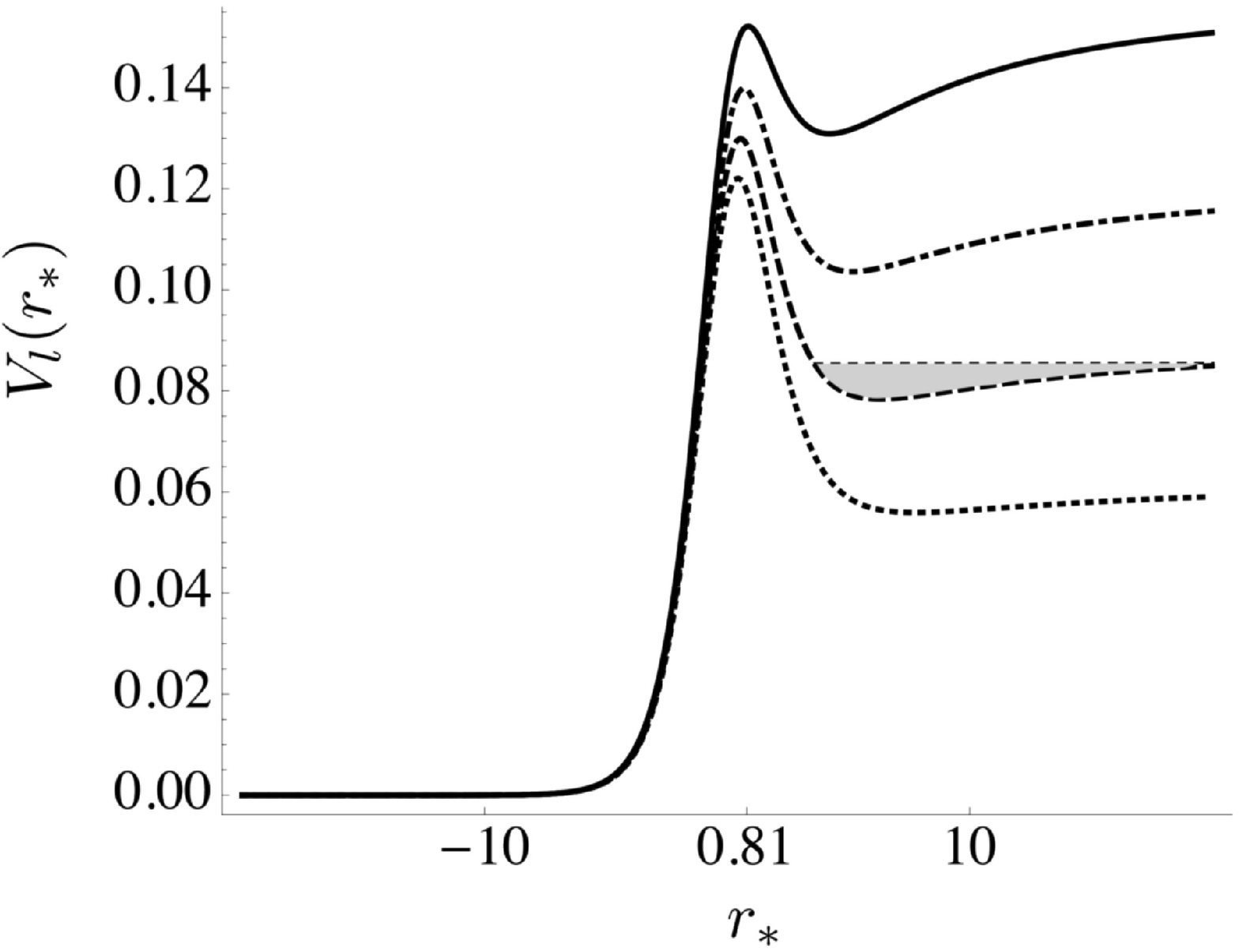}
\hfill
\includegraphics[width=0.49\textwidth]{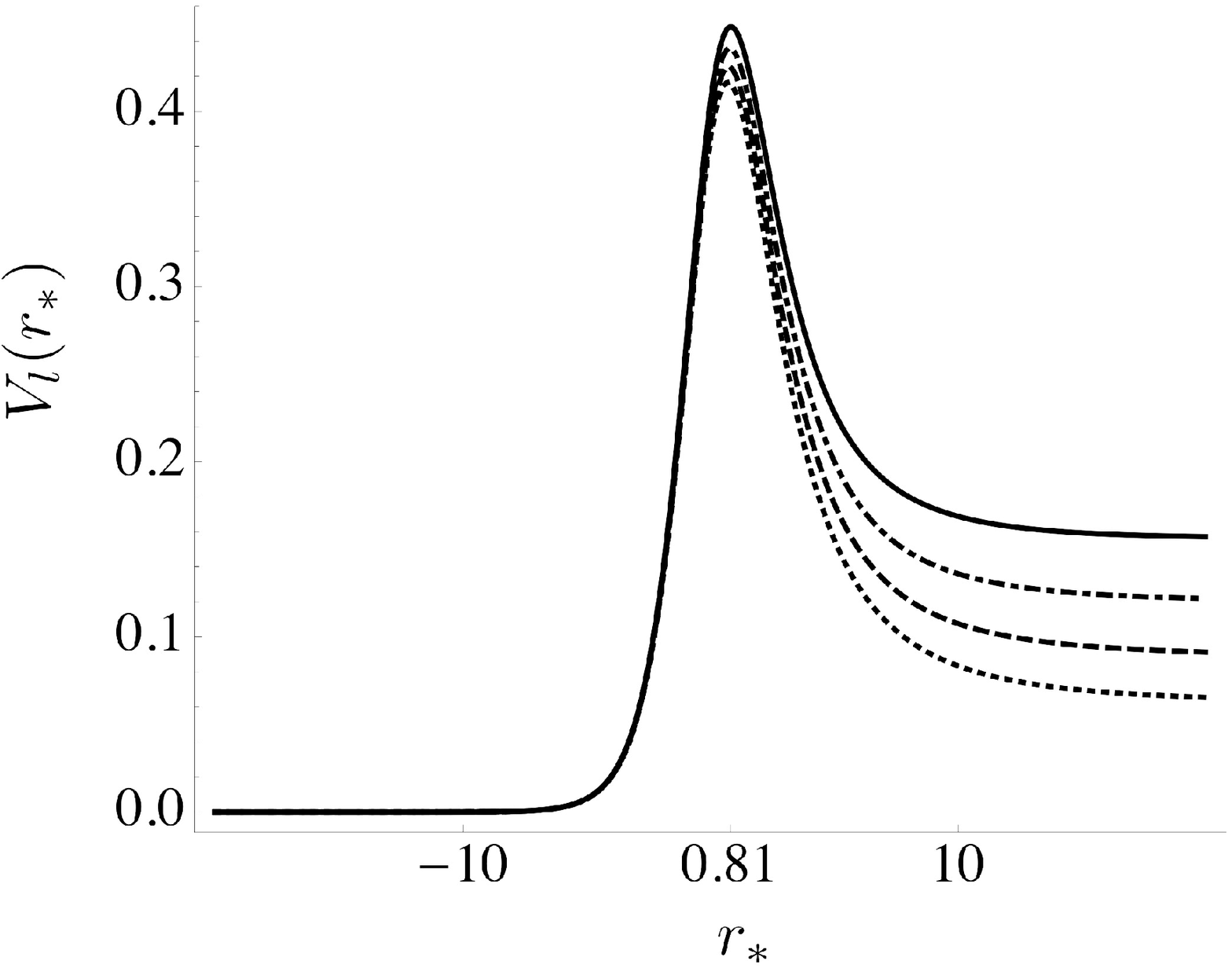}
\end{center}
\caption{\label{fig:potential} The potential as a function of the tortoise coordinate $r_*$: the lowest mode $l=s=0$ (left) and the first excited mode $l=1$ (right) for $m = 0.25$ (dotted), $m= 0.3$ (dashed), $m=0.35$ (dash-dotted) and $m=0.4$ (solid).}
\end{figure}

In this section we complement the recent work in Ref. \refcite{PhilippPerlick2015} and give the solutions of the perturbation equation for a massive scalar field on Schwarzschild background spacetime in EF coordinates. The propagation of such a massive scalar field is described by the Klein-Gordon equation with a mass term
\begin{align}
\left( \Box - m^2 \right) \Phi \equiv \left( \nabla^\mu \nabla_\mu - m^2 \right) \Phi = \left[1/\sqrt{-g} \dfrac{\partial}{\partial x^\mu}\left( \sqrt{-g} ~ g^{\mu\nu} \dfrac{\partial}{\partial x^\nu} \right) - m^2 \right] \Phi = 0 \, , \label{eq:KleinGordon}
\end{align}
where $m$ is the mass of the scalar boson. The scalar Klein-Gordon field $\Phi$ is a function of all the coordinates $t,r,\vartheta,\varphi$ and can be expanded into scalar spherical harmonics $Y_{lm}(\vartheta,\varphi)$ as shown in many standard textbooks, see e.g. Ref. \refcite{FrolovNovikovBook},
\begin{align}
\Phi = \Phi(t,r_*,\vartheta,\varphi) = \sum^{\infty}_{l=0} \sum^{l}_{m= -l} \dfrac{\Psi_l(t,r_*)}{r} Y_{lm}(\vartheta,\varphi) \, , \label{eq:KG_expansion}
\end{align}
where we introduce the tortoise coordinate 
\begin{align}
r_* &= r+\log(r-1) \, .
\end{align}
With the ansatz \eqref{eq:KG_expansion} we obtain the wave equation
\begin{align}
\left[ \diffp[2]{}{r_*} - \diffp[2]{}{t} - V_{l}(r) \right] \Psi_{l}(t,r_*) = 0 \, , \quad V_{l}(r) = \dfrac{(r-1)}{r} \left( \dfrac{l(l+1)}{r^2} + \dfrac{1}{r^3} + m^2 \right) \, . \label{eq:massive_waveeqn}
\end{align}
This equation is sometimes called the time-dependent Regge-Wheeler equation for a massive spin $s=0$ field. In their original work Regge and Wheeler \cite{ReggeWheeler1957} derived the analogous equation for the massless $s=2$ case.
The lowest mode ($l=s=0$) and the first excited mode $l=s+1=1$ of the  potential $V_{l}(r)$ are shown in Fig. (\ref{fig:potential}) for different values of the mass $m$. The maxima are located near the light-sphere radius $r=3/2$. In the limit $r \to \infty$ the value of the potential is given by the squared  mass. This property distinguishes the present situation from the massless case which is obtained in the limit $m \to 0$. A detailed numerical analysis of (quasi-) resonant states (i.e., the gray region shown in Fig.~\ref{fig:potential}) in this potential can be found in Ref. \refcite{Barranco2011}, where Schwarzschild coordinates were used in the discussion but EF coordinates were considered for the numerical evolution as well.

\begin{figure}
\begin{center}
\includegraphics[width=0.49\textwidth]{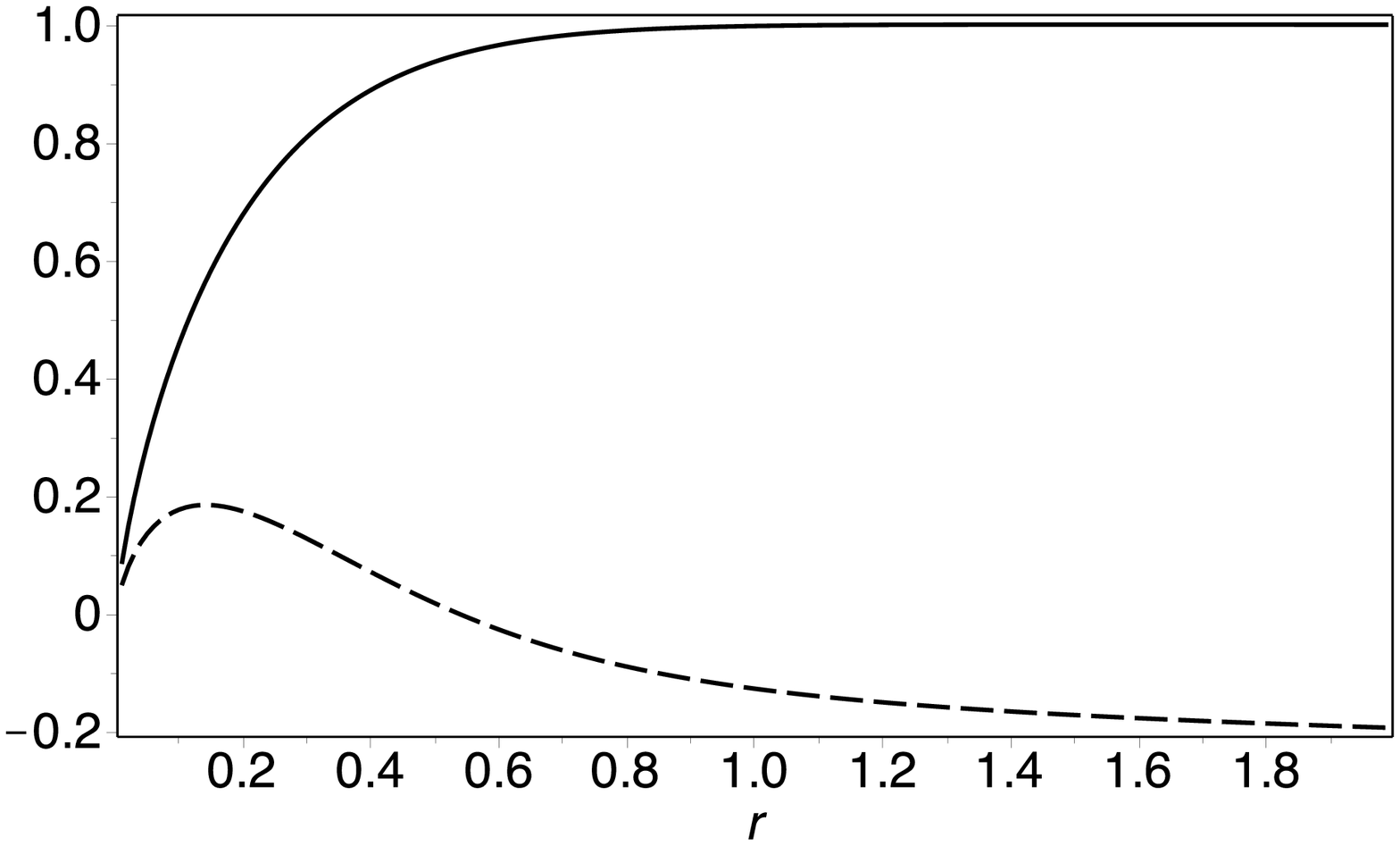}
\hfill
\includegraphics[width=0.49\textwidth]{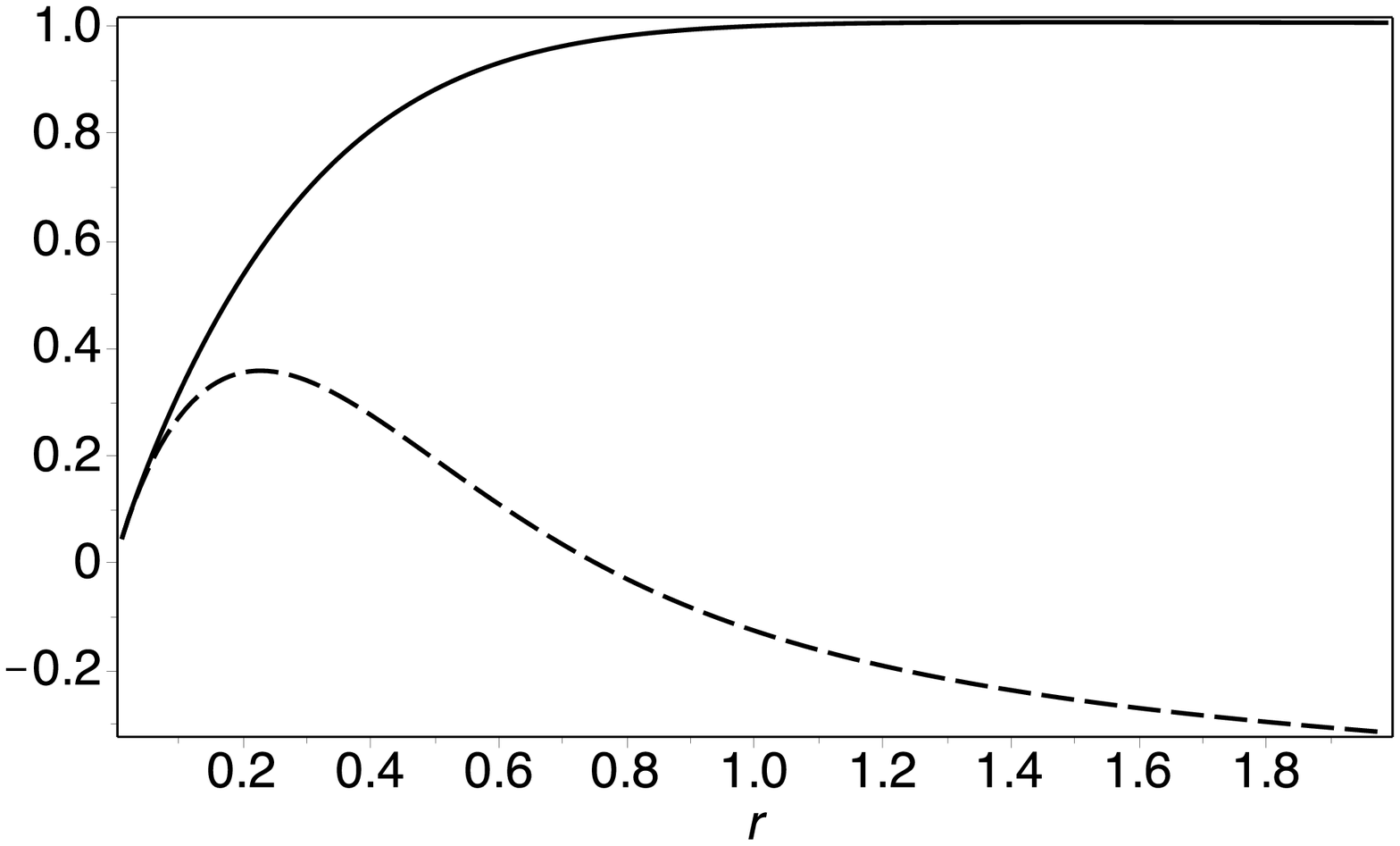} \\
\includegraphics[width=0.49\textwidth]{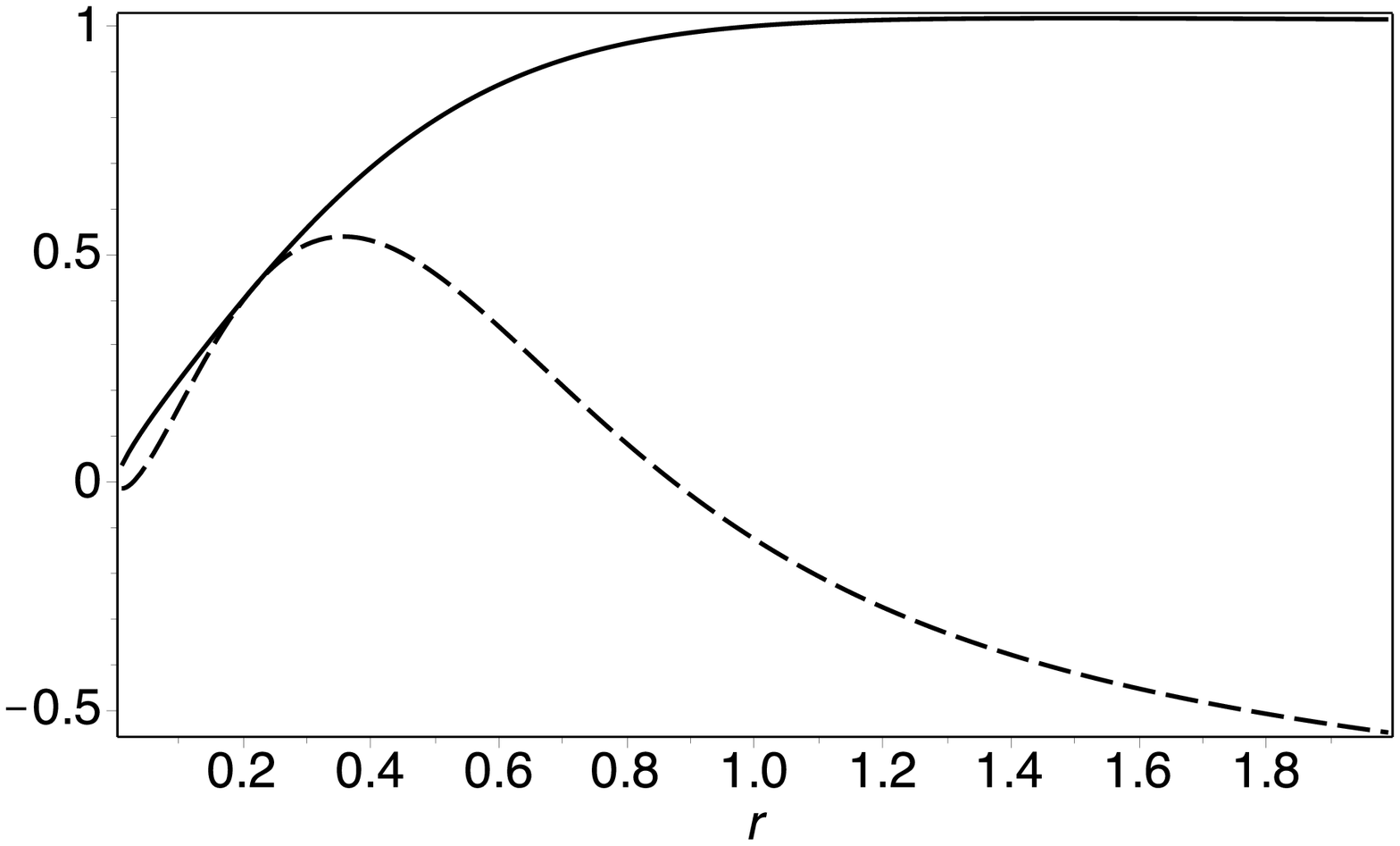}
\hfill
\includegraphics[width=0.49\textwidth]{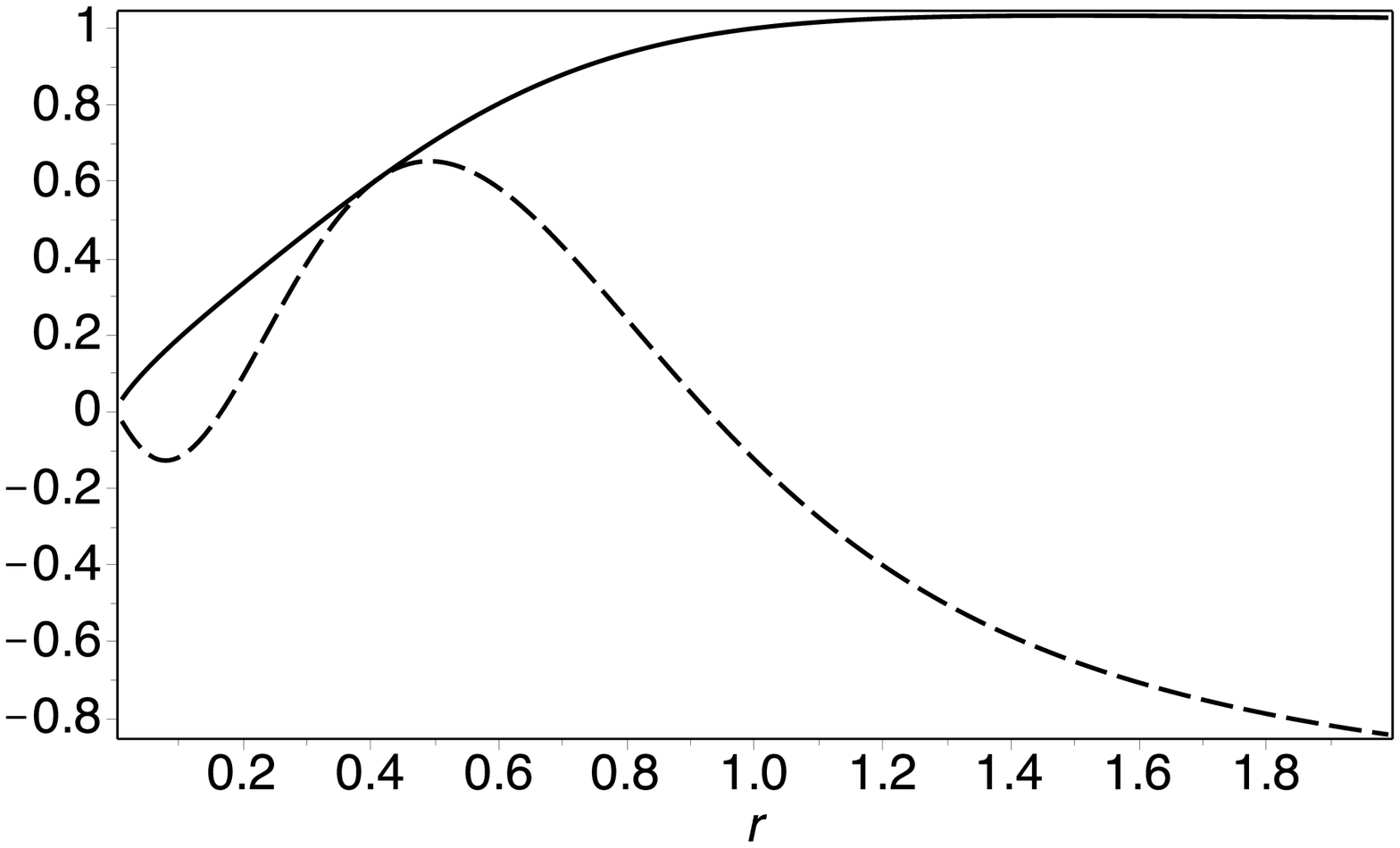} \\
\includegraphics[width=0.49\textwidth]{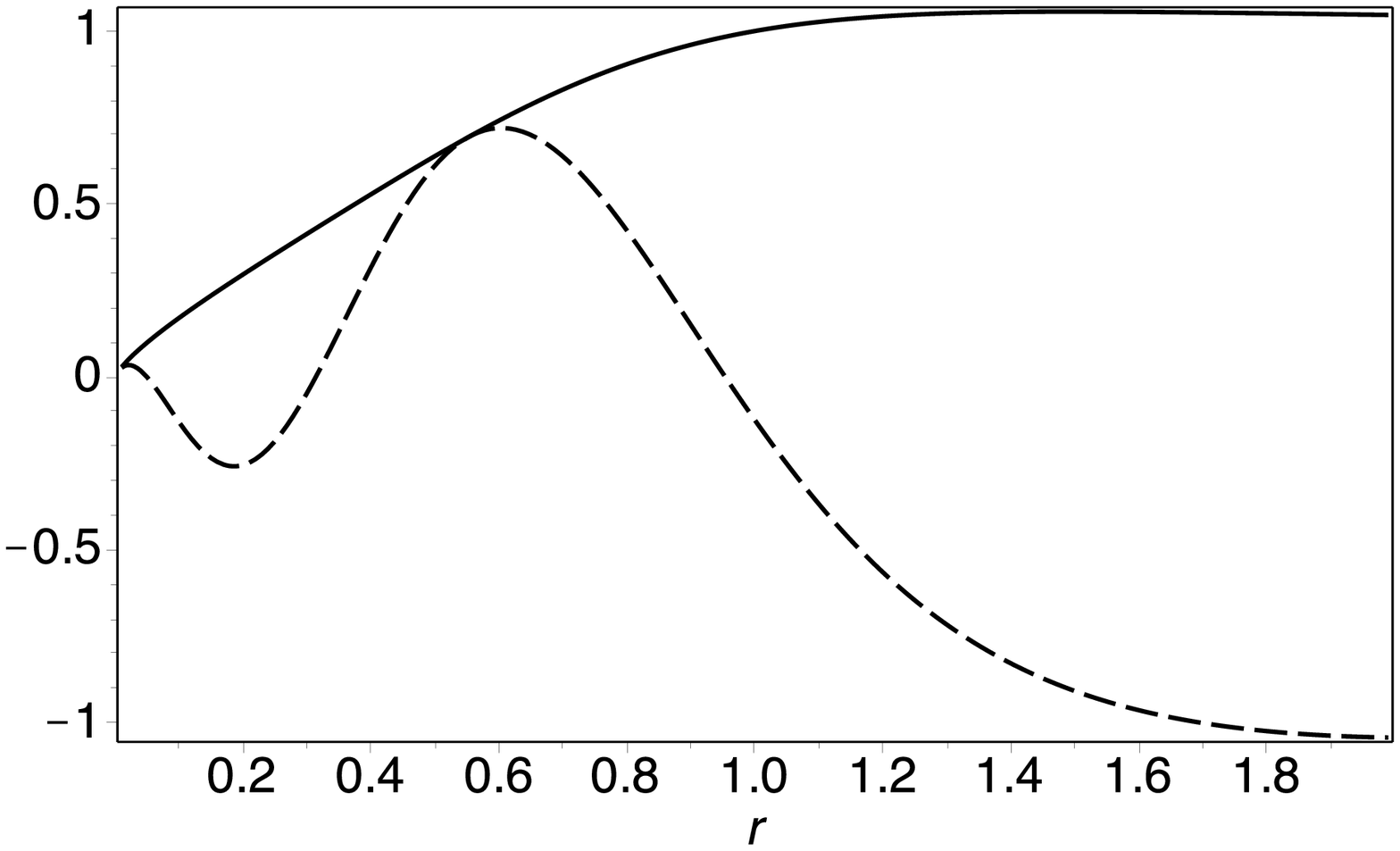}
\hfill
\includegraphics[width=0.49\textwidth]{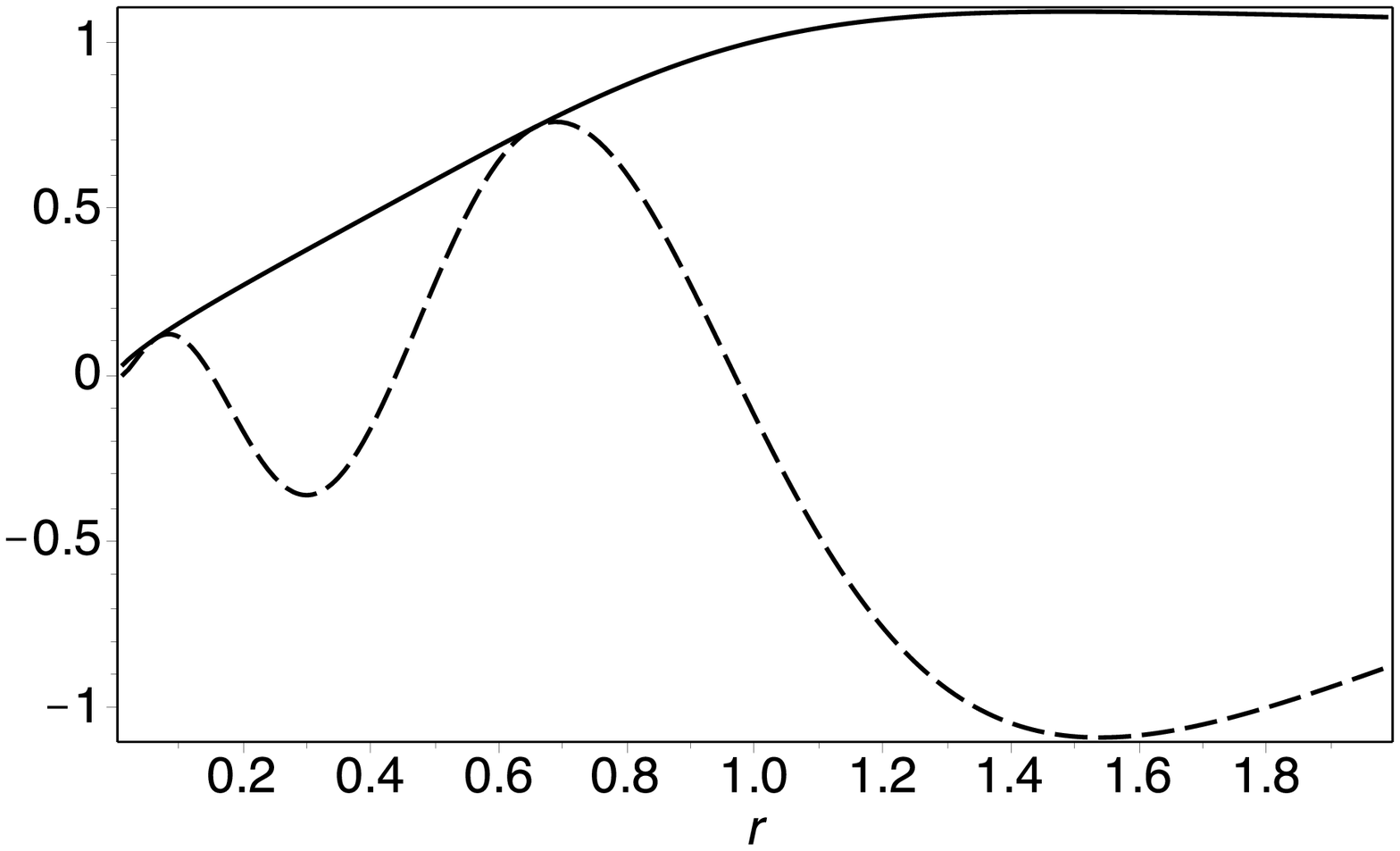}
\end{center}
\caption{\label{fig:Frobenius1_EF} The regular solution $R^{I}_{\omega l}(r;1)$ for a massive scalar field in EF coordinates for $l=0,1$ (top), $l=2,3$ (middle), $l=4,5$ (bottom) for the choice $\omega=4, \, m=0.4$. The modulus (solid) and the real part (dashed) are shown. The horizon of the Schwarzschild black hole is located at $r=1$.}
\end{figure}

Using the results of Ref. \refcite{PhilippPerlick2015} we can transform the wave equation \eqref{eq:massive_waveeqn} into a wave equation in Eddington-Finkelstein coordinates that allows us to obtain analytical and regular solutions around the black hole horizon at $r=1$.  There is no known exact analytic solution for the wave equation in the form \eqref{eq:massive_waveeqn} due to the involved potential term.
The corresponding wave equation in EF coordinates is \cite{PhilippPerlick2015}
\begin{align}
\left[ \left(\dfrac{r-1}{r}\right)^2 \diffp[2]{}{r} + 2  \dfrac{(r-1)}{r} \diffp{}{r v} + \dfrac{(r-1)}{r^3} \diffp{}{r} -V_l(r) \right] \Psi_l(v,r) = 0 \, . \label{eq:massive_waveeqn_EF}
\end{align}
This equation can be further transformed into a radial equation using the ansatz
\begin{align}
\Psi(v,r) = \mathrm{e}^{-i\omega v} R_{\omega l}(r)
\end{align}
that separates the EF time coordinate $v$. The radial equation that the function $R_{\omega l}(r)$ has to fulfill is then given by
\begin{align}
\left[ \diff[2]{}{r}+ \left( \dfrac{1-2i \omega r^2}{r(r-1)} \right) \diff{}{r} - \left( \dfrac{l(l+1)}{r(r-1)} + \dfrac{1}{r^2(r-1)} + \dfrac{m^2 r}{(r-1)}  \right) \right] R_{\omega l}(r) = 0 \, .
\end{align}
We recognize regular singularities at the origin $r=0$ and at the horizon $r=1$. Moreover, there is an irregular singularity at $z=\infty$ which can be seen by substituting $\zeta = 1/r$ and investigating the behavior at $\zeta=0$. With these properties the radial equation belongs to the class of singly confluent Heun equations.
The transformation
\begin{align}
R_{\omega l}(r) = r ~ \mathrm{e}^{\left( i \omega + \sqrt{m^2 - \omega^2} \right) r} H_{\omega l}(r)
\end{align}
yields a CHE in the form of \eqref{eq:CHE_MapleForm} with the parameters
\begin{align}
a = 2 \sqrt{m^2-\omega^2} \, , \quad b = 0 \, , \quad c =-2i\omega \, , \quad \mu = i\omega + l(l+1) + \sqrt{m^2-\omega^2} \, ,\notag \\
\nu = -2i \sqrt{m^2-\omega^2} \omega -l(l+1) - m^2 + \sqrt{m^2-\omega^2} + 2\omega^2 - i\omega \, .
\end{align}
Hence, the GRS for the radial equation becomes, using the general result of Ref. \refcite{PhilippPerlick2015},
\begin{align}
\begin{pmatrix}
1 & 1 & 2 &  \\ 
0 & 1 & \infty & ;r \\[5pt]
0 & 0 & \left(1-2i\omega + \dfrac{2\omega^2 - m^2}{2 \sqrt{m^2-\omega^2}} \right) &  \\[10pt]
0 & 2i \omega & \left( 1 - \dfrac{2\omega^2 - m^2}{2 \sqrt{m^2-\omega^2}} \right)  &  \\[10pt]
 &  & 0 &  \\[2pt]
 &  & \left(-2\sqrt{m^2-\omega^2}\right) & 
\end{pmatrix} \, .
\label{eq:EF_GRS}
\end{align}
We can give two independent solutions $R^{I,II}_{\omega l}(r;1)$ at the regular singularity at the horizon ($r=1$) in terms of the standard confluent Heun function
\begin{subequations}
\label{eq:EF_Frobius}
\begin{align}
R^{I}_{\omega l}(r;1) &= r ~ \mathrm{e}^{\left( i \omega + \sqrt{m^2 - \omega^2} \right) r} \notag\\
&\times \mathrm{HeunC}(-2\sqrt{m^2-\omega^2},-2i\omega,0,m^2-2\omega^2,-l(l+1)+2\omega^2-m^2,1-r) \, ,\\
R^{II}_{\omega l}(r;1) &= r ~ \mathrm{e}^{\left( i \omega + \sqrt{m^2 - \omega^2} \right) r} (r-1)^{2i\omega} \notag\\
&\times \mathrm{HeunC}(-2\sqrt{m^2-\omega^2},2i\omega,0,m^2-2\omega^2,-l(l+1)+2\omega^2-m^2,1-r) \, .
\end{align}
\end{subequations}
From this we recover indeed the previous solutions of the massless case \cite{PhilippPerlick2015} in the limit $m\to 0$. While the first solution $R^{I}_{\omega l}(r;1)$ is regular at the black hole horizon and corresponds to purely ingoing radiation at that radius, the second solution $R^{II}_{\omega l}(r;1)$ is not regular there since the phase performs infinitely many turns on the unit circle in the complex plane due to the term $(r-1)^{2i\omega}$ that has no well defined limit for $r \to 1$. When we want to imply causal boundary conditions, i.e., no radiation should escape the black hole horizon, we only need to consider the first, regular, solution. Figure (\ref{fig:Frobenius1_EF}) shows the $l=0$ to $l=5$ modes of this regular solution $R^{I}_{\omega l}(r;1)$.

The local solutions \eqref{eq:EF_Frobius} are, unfortunately, only useful in the rather small domain of convergence $|r-1| < 1~$  but can be analytically continued to the entire domain $r \in \, ]0,\infty[$ by the procedure that was demonstrated in Ref. \refcite{PhilippPerlick2015} for the massless case and a general bosonic spin $s$ in EF coordinates. We will sketch how such a continuation procedure works for the case of massless fields in PG coordinates in the next section.
\section{\label{sec:PG} Massless fields in Painlev\'{e}-Gullstrand coordinates}
We start now with a brief reconsideration of the results derived in Ref. \refcite{PhilippPerlick2015} that will be needed for this section. In that article we have shown that the wave equation on Schwarzschild background that describes the propagation of massless scalar ($s=0$), electromagnetic ($s=1$) or gravitational ($s=2$) perturbations written in PG coordinates is given by the partial differential equation
\begin{multline}
\left[ \left( \dfrac{r-1}{r} \right)^2 \diffp[2]{}{r} - \dfrac{(r-1)}{r} \diffp[2]{}{v}
+ 2 \dfrac{(r-1)}{r^{3/2}} \diffp{}{r v} \right. \\ \left. - \dfrac{(r-1)}{2 r^{5/2}}  \diffp{}{v}
 + \dfrac{(r-1)}{r^3} \diffp{}{r} - V_{s l}(r) \right] \Psi(v,r) = 0 \, ,\label{eq:waveEqn_PG}
\end{multline}
where the spin-dependent potential is given by
\begin{align}
V_{s l}(r) = \dfrac{(r-1)}{r^3} \left( l(l+1) + \dfrac{1-s^2}{r} \right) \, .
\end{align}
We can separate now the PG time coordinate $v$ using the separation ansatz
\begin{align}
\Psi(v,r) = \mathrm{e}^{-i\omega v} R_{\omega s l}(r) \, .
\label{eq:ansatzPG}
\end{align}
Thereby we introduce the frequency $\omega$ of the perturbation that has the dimension of an inverse length in our units; so the combination $\omega M$ is a dimensionless quantity. The harmonic time dependence in the ansatz \eqref{eq:ansatzPG} leads to a purely radial perturbation equation, that is
\begin{multline}
\left[ r(r-1) \diff[2]{}{r} + \left( 1- 2i \omega r^{3/2} \right) \diff{}{r} \right. \\
\left. + \left( r^2 \omega^2 + \dfrac{i\omega r^{1/2}}{2} - l(l+1) - \dfrac{1-s^2}{r} \right) \right] R_{\omega s l}(r) = 0 \, .\label{eq:radialEqn_PG}
\end{multline}
This radial equation is a second order, ordinary, linear and homogeneous differential equation. We further recognize that equation \eqref{eq:radialEqn_PG} possesses regular singularities at $r=0,1$. Using the transformation $\zeta =1/r$ we are able to find an irregular singularity at $\zeta = 0$, which correspond to $r=\infty$. With these singularities the radial equation could belong to the class of singly confluent Heun equations \cite{Heun1888, SlavyanovLayBook}. Indeed,  a transformation that maps \eqref{eq:radialEqn_PG} into the Maple-form of the confluent Heun equation is given by
\begin{align}
R_{\omega s l}(r) = r^{s+1} (\sqrt{r}-1)^{i \omega} \left( r^{3/2} - \sqrt{r} + r-1 \right)^{-i \omega} \mathrm{e}^{i\omega (2\sqrt{r}+r)} H_{\omega s l} (r) \, .\label{eq:PG_Maple_trafo}
\end{align}
The function $H_{\omega s l}(r)$ has to fulfill the differential equation \eqref{eq:CHE_MapleForm} with the parameters
\begin{align}
a &= 2i \omega \, , \quad b= 2s \, , \quad c = -2i\omega \, , \quad \mu = 4i\omega s + 2i\omega +l(l+1) - s(s+1) \, , \notag \\
\nu &= -2i\omega s + 4\omega^2 +s(s+1) -l(l+1) \, .
\end{align}
Hence, the GRS for this equation can be given by using the general result \cite{PhilippPerlick2015}, which yields
\begin{align}
\begin{pmatrix}
1 & 1 & 2 &  \\ 
0 & 1 & \infty & ;r \\ 
0 & 0 & s+1-2i\omega &  \\ 
-2s & 2i \omega & s+1 &  \\ 
 &  & 0 &  \\ 
 &  & -2i\omega & 
\end{pmatrix} \, .\label{eq:PG_GRS}
\end{align}
\subsection{Local solutions at the horizon}
\begin{figure}[t]
\begin{center}
\includegraphics[width=0.49\textwidth]{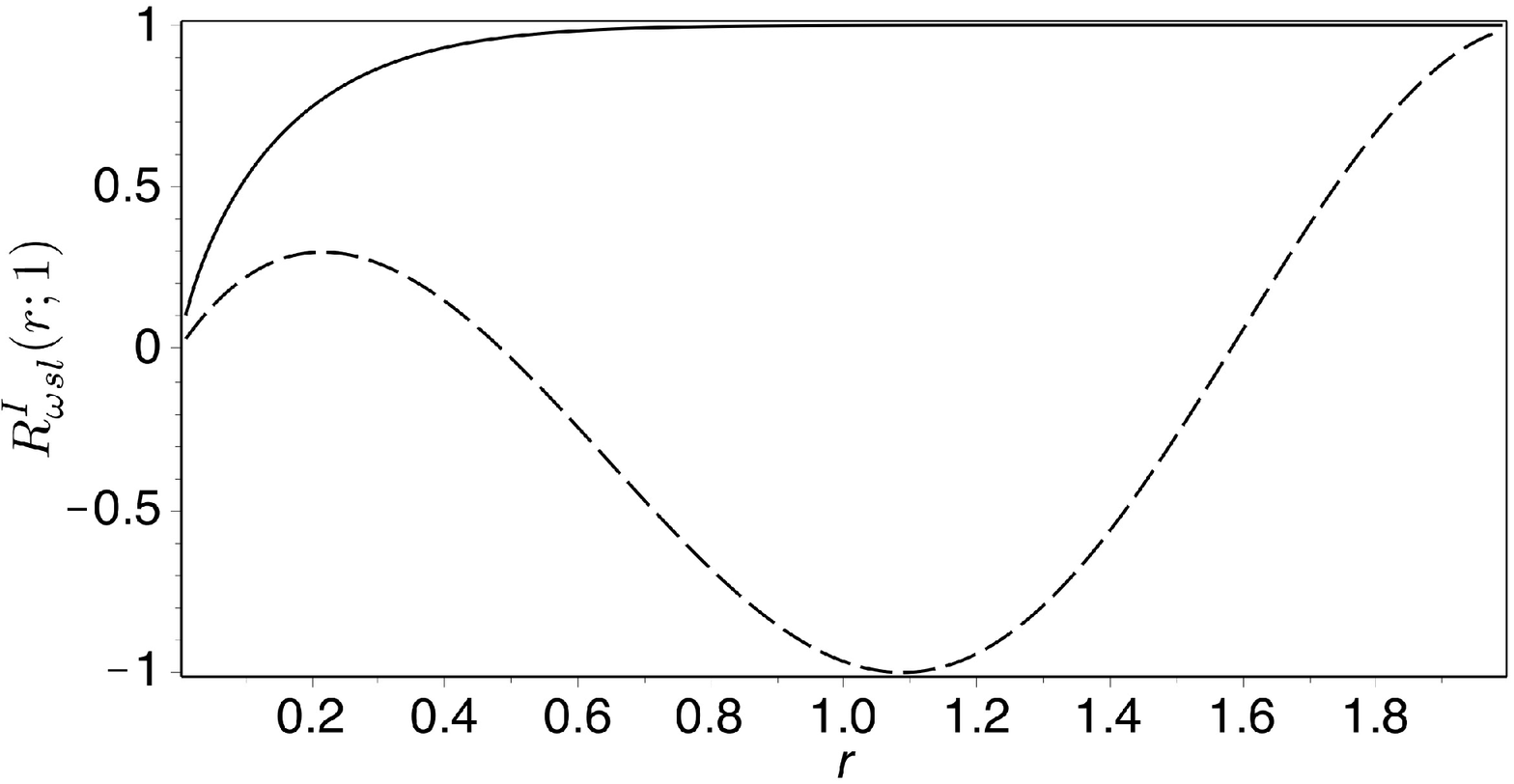}
\hfill
\includegraphics[width=0.49\textwidth]{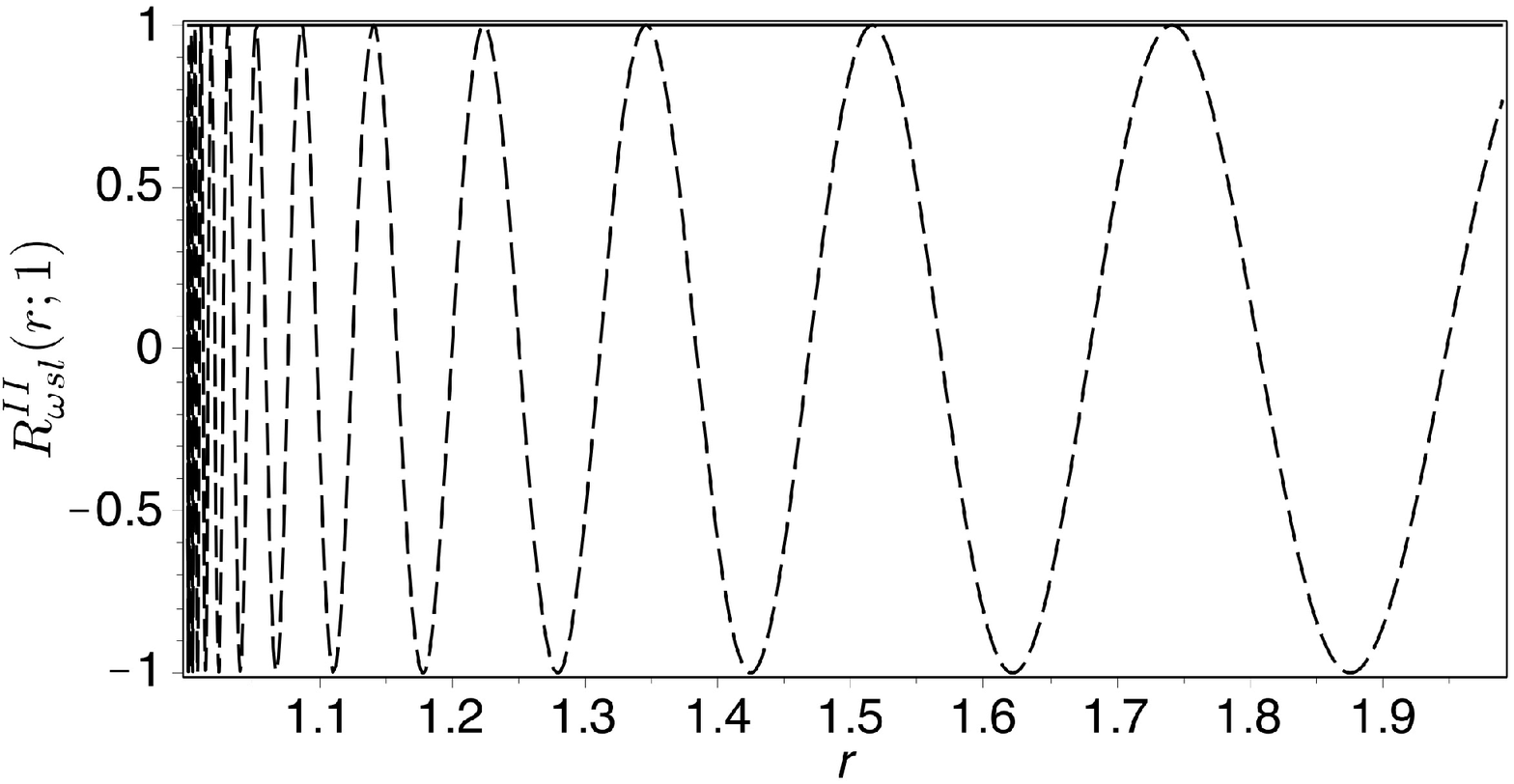}
\end{center}
\caption{\label{fig:PG_Frobenius} The two local solutions around the horizon: the regular solution $R^{I}_{\omega s l}(r;1)$ (left) and the second solution $R^{II}_{\omega s l}(r;1)$ (right) for $\omega = 6,\, l=s=0$. The modulus (solid) and real part (dashed) are shown and the horizon is located at $r=1$.}
\end{figure}

We can construct two independent solutions at the regular singularity at the horizon ($r=1$) using the confluent Heun function \cite{PhilippPerlick2015, MapleHeun}. These two solutions are
\begin{subequations}
\label{eq:PG_Frobenius}
\begin{align}
R^I_{\omega s l}(r;1) &= r^{s+1} (\sqrt{r}-1)^{i \omega} \left( r^{3/2} - \sqrt{r} + r-1 \right)^{-i \omega} \mathrm{e}^{i\omega (2\sqrt{r}+r)} \notag \\
&\times \mathrm{HeunC}(-2i\omega, -2i\omega, 2s,-2\omega^2,-l(l+1)+s^2+2\omega^2,1-r) \, ,\\
R^{II}_{\omega s l}(r;1) &=r^{s+1} (\sqrt{r}-1)^{i \omega} \left( r^{3/2} - \sqrt{r} + r-1 \right)^{-i \omega} \mathrm{e}^{i\omega (2\sqrt{r}+r)} \notag \\
&\times (r-1)^{2i\omega} \mathrm{HeunC}(-2i\omega, 2i\omega, 2s,-2\omega^2,-l(l+1)+s^2+2\omega^2,1-r) \, .
\end{align}
\end{subequations}
The first solution $R^{I}_{\omega s l}(r;1)$ describes a wave that propagates into the black hole horizon and is totally regular when approaching this radius. The second solution $R^{II}_{\omega s l}(r;1)$ on the other hand describes outgoing radiation at $r=1$ and is not well defined when approaching the horizon since the phase of this solution performs infinitely many turns on the unit circle in the complex plane in the limit $r \to 1$, as we have seen in section \ref{sec:massive} before. Figure (\ref{fig:PG_Frobenius}) shows the real part and the modulus of both solutions and clearly demonstrates the corresponding behavior at the horizon.
\subsection{Asymptotic series at spatial infinity}
Using the entries of the third column of the GRS \eqref{eq:PG_GRS} we can construct the Thom\'{e}-type solutions (asymptotic series) at the irregular singularity at spatial infinity. These solutions are \cite{SlavyanovLayBook,PhilippPerlick2015}
\begin{subequations}
\label{eq:PG_asymptotic}
\begin{align}
R^{I}_{\omega s l}(r;\infty) &= r^{s+1} (\sqrt{r}-1)^{i \omega} \left( r^{3/2} - \sqrt{r} + r-1 \right)^{-i \omega} \notag \\
&\times \mathrm{e}^{i\omega (2\sqrt{r}+r)}  \sum_{k=0}^{\infty} \rho_k r^{-(k+s+1-2i\omega)} \, ,\\
R^{II}_{\omega s l}(r;\infty) &= r^{s+1} (\sqrt{r}-1)^{i \omega} \left( r^{3/2} - \sqrt{r} + r-1 \right)^{-i \omega} \notag \\
&\times \mathrm{e}^{i\omega (2\sqrt{r}+r)}  \mathrm{e}^{-2i\omega r} \sum_{k=0}^{\infty} \sigma_k r^{-(k+s+1)} \, ,
\end{align}
\end{subequations}
and the series expansion coefficients $\rho_k$ and $\sigma_k$ can be calculated in terms of recurrence relations that are obtained by inserting the solutions above into the differential equation.
The first solution, together with the separated time dependence $\sim \mathrm{e}^{-i\omega v}$ corresponds to radiation that is propagating into the singularity at $r=\infty$ (outgoing), while the second solution describes waves that emerge from this singularity and propagate to smaller radii (ingoing). 
% The scheme in figure \ref{fig:InOut} sketches these in- and outgoing properties for all the Frobenius solutions and asymptotic series.
%\begin{figure}[]
%\begin{center}
%\includegraphics[width=0.49\textwidth]{InOut.pdf}
%\end{center}
%\caption{\label{fig:InOut} In- and outgoing properties for the local Frobenius solutions at the %horizon $R^{I,II}_{\omega s l}(r;1)$ and the asymptotic series at spatial infinity %$R^{I,II}_{\omega s l}(r;\infty)$.}
%\end{figure}
\subsection{Analytic continuation of Frobenius solutions}
Both local Frobenius solutions at the horizon $R^{I,II}_{\omega s l}(r;1)$ are convergent within the unit circle centered at $r=1$ in the complex plane as explained in Ref. \refcite{PhilippPerlick2015}. However, it is possible to perform an analytic continuation that combines the Frobenius solutions \eqref{eq:PG_Frobenius} with the asymptotic series \eqref{eq:PG_asymptotic} and in this way to obtain solutions on the entire open interval $r \in \, ]0, \infty[$ . This procedure can be partly traced back to work of Jaff\'{e} \cite{Jaffe1934} and is demonstrated in detail for the radial perturbation equation for massless fields in EF coordinates in Ref. \refcite{PhilippPerlick2015} using the general formalism introduced by Slavyanov and Lay \cite{SlavyanovLayBook} in terms of the central two point connection problem (CTCP). We refer the reader to these works and restrict ourselves here to a brief sketch of how to apply this procedure to the present case of PG coordinates. The five-step procedure that we apply is the following:
\subsubsection*{1. Shift of singularities and reordering the differential equation}
We perform the simple transformation $z=r-1$ to shift the involved regular singularity from $r=1$ to $z=0$ and reorder the resulting differential equation in the form
\begin{align}
\left[ \diff[2]{}{z} + \left( \dfrac{A_2}{z+z_*} + \dfrac{A_1}{z} + G_0 \right) \diff{}{z} + \left( \dfrac{C_2}{z+z_*} + \dfrac{C_1}{z} \right) \right] y_{\omega s l}(z) = 0
\end{align} 
with the parameter
\begin{align}
z_* &= 1 \, , \quad G_0 = 2i \omega \, , \quad A_1 = 1-2i\omega \, , \quad A_2 = 1+2s \, , \notag \\
C_1 &= -2i\omega s + 4 \omega^2 + s(s+1) - l(l+1)\, , \quad C_2 =4i\omega s  + 2i\omega + l(l+1) - s(s+1) \, ,  \label{eq:PG_CTCP_parameter}
\end{align}
to match the general formulas given in Ref. \refcite{SlavyanovLayBook}. Note that the irregular singularity at $r=\infty$ remains at $z=\infty$.
\subsubsection*{2. s-homotopic transformation of the dependent variable} 
\begin{figure}[]
\begin{center}
\includegraphics[width=0.8\textwidth]{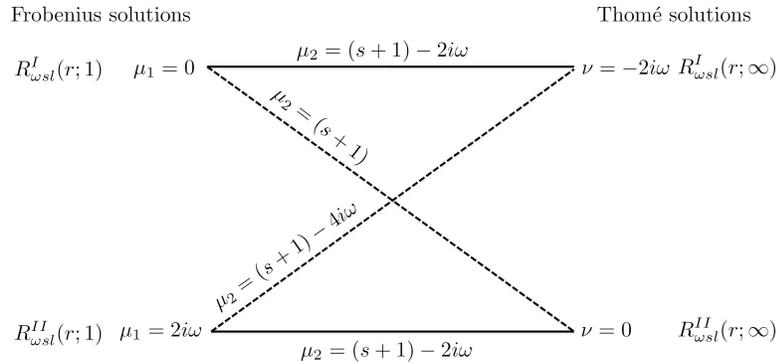}
\end{center}
\caption{\label{fig:exponents} A scheme showing the proper choice of the exponents $\nu, \mu_1$ and $\mu_2$ related to local solutions for the connection problem.}
\end{figure}

The next step is to perform a special transformation of the dependent variable $H_{\omega s l}(z)$. The transformation we need is a so-called s-homotopic transformation and given by
\begin{align}
&H_{\omega s l}(z) \mapsto \Xi_{\omega s l}(z) \notag \\
&H_{\omega s l}(z) = \mathrm{e}^{\nu z} \, z^{\mu_1} \, (z+1)^{\mu_2} \, \Xi_{\omega s l}(z) \, .
\end{align}
The three new exponents $\nu, \mu_1, \mu_2$ are connected to characteristic exponents at the two involved singularities. Similar to the treatment in Ref. \refcite{PhilippPerlick2015} we obtain the possible values
\begin{itemize}
\setlength{\parskip}{1pt}
\item[-] $\mu_1$ is the indicial exponent of the chosen Frobenius solution at $z = 0$. $\Rightarrow \mu_1 = 0$ or $\mu_1 = 2i \omega$ according to the GRS \eqref{eq:PG_GRS}.
\item[-] $\nu$ is the characteristic Thom\'{e} exponent of order one and depends on the chosen Thom\'{e} solution at $z=\infty$. The possibilities are $\nu = 0$ or $\nu = -2i \omega$.
\item[-] For large $z$ we get $z^{\mu_1} (z+1)^{\mu_2} \approx z^{\mu_1+\mu_2}$. Thus,  $\mu_1 + \mu_2$ is the characteristic exponent of order zero of the chosen Thom\'{e} solution. $\Rightarrow \mu_2 = (s+1)$, $\mu_2 = (s+1)- 2i \omega$ or $\mu_2 = (s+1) -4i\omega$.
\end{itemize}
As a summary of this step we sketch in Figure (\ref{fig:exponents}) the appropriate choice of the three exponents according to the behavior at the singularities that is determined by the Frobenius solutions and the asymptotic series.
\subsubsection*{3. M\"obius transformation of the independent variable} 
We apply now a transformation of the independent variable. The necessary transformation in this step is given by the M\"obius transformation
\begin{align}
z \mapsto x = \dfrac{z}{z+1}
\end{align}
that maps the irregular singularity from $z=\infty$ to $x=1$ at the boundary of the unit circle. The other singularity involved is now located at $x=0$ and we have mapped the infinite interval $z \in \,[0,\infty]$ to the finite interval of unit length $x \in \, [0,1]$.
\subsubsection*{4. Series expansion ansatz} 
Now, we solve the resulting differential equation for $\Xi_{\omega s l}(x)$ with the series expansion ansatz
\begin{align}
\Xi_{\omega s l}(x) = \sum_{k=0}^\infty \xi_k(\omega,s,l) ~ x^k \, . \label{eq:CTCP_ansatz}
\end{align}
This yields a three term recurrence relation for the series expansion coefficients $\xi_k(\omega, s, l)$ in the form
\begin{align}
\xi_0 &= \text{arbitrary} \notag \\
\alpha_0 \xi_1 + \beta_0 \xi_0 &= 0 \notag \\
\alpha_k \xi_{k+1} + \beta_k \xi_k + \gamma_k \xi_{k-1} &= 0 \, ,  \label{eq:CTCP_recursion}
\end{align}
where the $\alpha_k\, , \beta_k$ and $\gamma_k$ depend on the parameters \eqref{eq:PG_CTCP_parameter} and, therefore, also on $\omega, s$ and $l$. The actual form of these quantities can be derived using the results given in Ref. \refcite{PhilippPerlick2015}. The series representation of the function $\Xi_{\omega s l}(x)$ is absolutely convergent on the interval $|x| < 1$\cite{Leaver1986,PhilippPerlick2015}.
\subsubsection*{5. Full solution of the radial equation} 
To construct the total solution to the radial equation in PG coordinates \eqref{eq:radialEqn_PG} we have to recover the radius variable $r$ that we started with. This is done by using the inverse M\"obius transformation
\begin{align}
x = \dfrac{z}{z+1} = \dfrac{r-1}{r} \, .
\end{align}
Now, we construct the total solution using again the transformation \eqref{eq:PG_Maple_trafo}. Hence, we obtain
\begin{multline}
R_{\omega s l}(r) = r^{s+1} (\sqrt{r}-1)^{i \omega} \left( r^{3/2} - \sqrt{r} + r-1 \right)^{-i \omega} \mathrm{e}^{i\omega (2\sqrt{r}+r)} 
\\
\times \mathrm{e}^{\nu (r-1)} (r-1)^{\mu_1} r^{\mu_2} \sum_{k=0}^\infty \xi_k(\omega,s,l) \dfrac{(r-1)^k}{r^k} \, ,\label{eq:CTCP_solution}
\end{multline}
where the exponents $\mu_{1,2}$ and $\nu$ must be chosen according to the physical boundary conditions with the help of Figure (\ref{fig:exponents}) and the $\xi_k$ follow from the mentioned three-term recurrence relation.
The solution \eqref{eq:CTCP_solution} is now absolutely convergent in the interval $r \in \, ]0.5,\infty[$ \cite{PhilippPerlick2015} and we have, thus,  solutions on the the entire interval $r \in \, ]0,\infty[$ since for the region $r \in \, ]0,2[$ the Frobenius solutions \eqref{eq:PG_Frobenius} in terms of the standard confluent Heun function can be used and the analytically continued solutions coincide with these local Frobenius solutions in the overlapping region.
\section{Conclusion}
We have shown that the propagation equations for bosonic spin $s$ fields on Schwarzschild background spacetime can be reduced to purely radial differential equations for Eddington-Finkelstein and Painlev\'{e}-Gullstrand coordinates. In these coordinate systems the radial equations admit (local) solutions in terms of convergent series. We complemented our recent work \cite{PhilippPerlick2015}, where massless fields were investigated in EF coordinates, by additionally analyzing a massive scalar field in EF coordinates and extending the treatment of the massless bosonic fields to the case of PG coordinates. In each of these cases it is possible to derive two local solutions around the regular singularity at the black hole horizon. One of these solutions is regular at $r=2M$ and connected to causal boundary conditions of purely ingoing radiation, and the other solution describes radiation emerging from the black hole. While the regular solution can be used to investigate scattering of radiation or to calculate quasi-normal modes, the second solution describes Hawking radiation and is, therefore, connected to black hole evaporation as shown in Ref. \refcite{PhilippPerlick2015}. We have further shown that using an analytic continuation procedure the bounded domain of convergence of the local solution (i.e., the unit circle around the horizon in the complex plane) can be expanded to the entire open interval between the origin and spatial infinity. It is a remarkable feature that the structure of the radial equations in EF and PG coordinates is the same. Both cases belong to the class of confluent Heun equations and can be solved using Frobenius series at the regular singularity at $r=2M$ that can analytically be continued up to, but not in general including, spatial infinity. The equations in PG coordinates are somewhat more complicated and we recommend, thus, to treat all these perturbations in EF coordinates. If, however, the situation should be linked to the measurement of a radially infalling observer on a timelike geodesic, one should use the solution we provided in Eqn. \eqref{eq:CTCP_solution}. This solution can be used to investigate various phenomena concerning massless bosonic waves on Schwarzschild background and can be adopted to a broad range of initial or boundary value conditions. The most interesting feature of the regular coordinate systems is the possibility to penetrate the horizon and thereby to  follow the radiation that falls into the black hole.
\section*{Acknowledgments}
The authors would like to thank Eugen Radu, Claus L\"ammerzahl, Domenico Giulini and Norman G\"urlebeck for valuable discussions during the preparation of the manuscript. Support from the Deutsche Forschungsgemeinschaft within the Research Training Group 1620 ``Models of Gravity" is gratefully aknowledged. The first author is financially supported by the German Collaborative Research Center 1128 ``geo-Q'' and acknowledges support from the Aveiro University in Portugal for accommodation during the VII Black Hole Workshop.

\bibliographystyle{ws-ijmpd}

\end{document}